\documentclass[11pt]{amsip-p-l}

\usepackage{amssymb}

\usepackage[vcentermath]{youngtab}
\usepackage{multirow}
\usepackage{slashed}

\def\half{\frac{1}{2}}
\def\tr{{\mathrm{tr}}}

\def\Z{\mathbb{Z}}
\def\F{\mathbb{F}}

\def\G{G}
\def\nhv{H-V}

\def\R{\mathbb{R}}

\def\P{\mathbb{P}}

\newtheorem{theorem}{Theorem}[section]

\newtheorem{conjecture}[theorem]{conjecture}

\theoremstyle{definition}

\theoremstyle{remark}

\numberwithin{equation}{section}

\begin{document}

\hspace*{\fill}MIT-CTP-4157
\vspace*{0.4in}

 \title[Six-dimensional quantum gravity]{Anomaly constraints and
string/F-theory geometry in 6D quantum gravity}


\author{Washington Taylor}
\address{Center for Theoretical Physics,\\
Department of Physics,\\
Massachusetts Institute of Technology;\\
77 Massachusetts Avenue,\\
Cambridge, MA 02139, USA}
\email{{\tt wati} {\rm at} {\tt mit.edu}}
\thanks{To appear in Proceedings of ``Perspectives in
Mathematics and Physics,'' conference in honor of I.\ M.\ Singer.
The author would like to thank collaborators Vijay Kumar and
David Morrison, with whom the work described in this paper was carried
  out.  The author is immensely grateful to IMS for regular
  stimulating discussions and interactions over the last two decades,
  and for his optimism and constant support over the years.
This research was
supported by the DOE under contract \#DE-FC02-94ER40818.
}


\date{\today}

\begin{abstract}
Quantum anomalies, determined by the Atiyah-Singer index theorem,
place strong constraints on the space of quantum gravity theories in
six dimensions with minimal supersymmetry.  The conjecture of ``string
universality'' states that all such theories which do not have
anomalies or other quantum inconsistencies are realized in string
theory.  This paper describes this conjecture and recent work by
Kumar, Morrison, and the author towards developing a global picture of
the space of consistent 6D supergravities and their realization in
string theory via F-theory constructions.  We focus on the discrete
data for each model associated with the gauge symmetry group and the
representation of this group on matter fields.  The 6D anomaly
structure determines an integral lattice for each gravity theory,
which is related to the geometry of an elliptically fibered Calabi-Yau
three-fold in an F-theory construction.  Possible exceptions to the
string universality conjecture suggest novel constraints on low-energy
gravity theories which may be identified from the structure of
F-theory geometry.
\end{abstract}

\maketitle

\section{Introduction}

String theory is not yet a mathematically well-defined system.
Nonetheless, the components of this theory which have been pieced
together so far suggest a framework in which quantum physics and
general relativity are unified.  In certain limits it seems that the
theory can be described geometrically and gives rise to 4D ``vacuum
solutions'', in which space-time has four macroscopic dimensions and
additional microscopic dimensions are curled up into a compact
manifold.  Each 4D string vacuum solution gives rise to a low-energy
physical theory in four dimensions which generally is a theory of
gravity coupled to Yang-Mills theory with matter fields transforming
under the Yang-Mills symmetry group.  The vast landscape of possible
four-dimensional string theory vacua presents a major challenge for
physicists who hope to derive predictions for observable physics from
string theory.  String theorists have struggled with this problem
since the early days of the subject, when it was realized that
ten-dimensional string theory could be compactified to four dimensions
on any Calabi-Yau threefold, giving a wide variety of low-energy
theories.  Efforts to remove unphysical massless scalar fields
associated with Calabi-Yau moduli, such as through inclusion of
generalized magnetic fluxes, have sharpened the problem further by
generating exponentially large (or infinite) classes of vacua
associated with each Calabi-Yau geometry (for a review of recent work
in this direction see {\it e.g.} \cite{flux-review}).  Beyond the
enormous range of string vacua which can be constructed using this and
other known mechanisms, it may be that even larger classes of string
vacua can be constructed using less well-understood non-geometric
compactifications.  Despite the wide range of possible string vacuum
constructions, the range of low-energy theories which exhibit no known
quantum inconsistencies when coupled to gravity seems to be far vaster
still.  The term ``swampland'' was coined by Vafa to characterize the
set of low-energy gravity theories with no known inconsistencies which
cannot be realized in string theory \cite{swampland}.  At present, the
swampland for four dimensions appears to be very large, and we lack
tools powerful enough to make strong global statements about the space
of consistent quantum gravity theories.

In six space-time dimensions, however, a global understanding of the
space of supersymmetric string vacua may be within reach.  In $4k+2$
dimensions, quantum anomalies can lead to a violation of
diffeomorphism or Yang-Mills invariance.  The condition that such
anomalies are absent strongly constrains the range of possible
quantum-consistent supergravity theories in six dimensions.  The
following conjecture was presented at the conference ``Perspectives in
Mathematics and Physics,'' in honor of I.\ M.\ Singer, and stated more
precisely in the paper \cite{universality}.

\begin{conjecture}[``String Universality in six dimensions'']
\label{c:conjecture}
All ${\mathcal N} = (1, 0)$ supersymmetric theories in 6D with one
gravity and one tensor multiplet which are free of anomalies or
other quantum inconsistencies admit a string construction.
\end{conjecture}
In the statement of this conjecture, ${\mathcal N} =(1, 0)$
supersymmetry is the minimal possible amount of supersymmetry in six
dimensions.  Supersymmetry is a graded symmetry relating bosonic
(commuting) to fermionic (noncommuting) fields.  The gravity and
tensor multiplets are combinations of fields transforming under
irreducible representations of the supersymmetry algebra.  We describe
some relevant features of these multiplets in more detail in the
following section.  Over the last year, further work with V.\ Kumar
\cite{finite} and with V.\ Kumar and D.\ Morrison \cite{KMT,
tensors} has led to a global picture of the space of constraints on
these low-energy supergravity theories, and an explicit correspondence
with the F-theory description of string vacua, not only for theories
with one tensor multiplet, but also for a much richer class of
theories with multiple tensor multiplets.  We have proven that when
the number of tensor multiplets is less than 9, there are only a
finite number of distinct gauge groups and matter representations
which can appear in consistent low-energy theories \cite{finite, tensors}.
This analysis has also, however, produced
some apparent counterexamples to the conjecture.  Either these
counterexamples represent new classes of string vacua, which cannot be
realized using conventional F-theory, or there must be additional
stringy constraints on the space of possible low-energy theories.  In
the latter case, if the conjecture is correct, it may be the case that
these constraints can be identified from the low-energy data
describing the theory as obstructions to a consistent quantum UV
completion.  
The analogue of Conjecture 1.1 in 10 dimensions was recently proven by
Adams, de Wolfe, and the author by demonstrating the quantum
inconsistency of the two models with gauge groups $U(1)^{496}$ 
and $E_8 \times U(1)^{248}$ which were
previously thought to lie in the swampland \cite{adt}.
Further progress towards proving or disproving the string
universality conjecture in 6 dimensions
will require achieving a better mathematical
or physical understanding of the set of possible string constructions
and low-energy constraints.

In this paper we describe some of the context and motivation for the
6D string universality
conjecture, and give a somewhat self-contained presentation of some of
the core mathematical arguments underlying recent progress in
understanding the relevant space of 6D supergravity theories.  The
group-theoretic structure of the anomaly cancellation conditions is
reviewed in Section \ref{sec:anomalies} and used in Section
\ref{sec:constraints} to place bounds on the set of gauge groups and
matter representations which can be realized in ${\mathcal N}= 1$
supersymmetric 6D theories of gravity coupled to Yang-Mills theory.
From the anomaly conditions, an integral lattice $\Lambda$ can be
associated with each such low-energy supergravity theory.  In Section
\ref{sec:F-theory}, we describe the connection with F-theory, where
$\Lambda$ is a sublattice of the lattice of integral second homology
classes on a complex surface which is the base of an elliptic
fibration.  In Section \ref{sec:examples} we give some explicit
examples of 6D theories and describe some apparent counterexamples to
the conjecture.  This leads to a discussion in the concluding section
\ref{sec:conclusions} of possible new constraints on low-energy
supergravity theories.

\section{6D supergravity: gauge groups, matter representations, and
anomalies}
\label{sec:anomalies}

In this section we describe some of the key mathematical structures in
6D supergravity theories.  In particular, we show how the set of
possible theories is constrained by using group theory to analyze
quantum anomaly constraints.  Our focus here is on topological, rather
than analytic aspects of these theories.  Thus, we are primarily
interested in classifying the discrete data associated with the set of
quantum fields in the theory, in particular the Yang-Mills symmetry
group and associated representations.  The following abbreviated
outline of the structure of 6D supergravity theories highlights this
discrete structure, and elides many of the details of the physics of
these theories.  For a more detailed description of 6D ${\mathcal
N} = (1, 0)$ supergravity theories, see the original papers
\cite{Nishino-Sezgin, Romans, GSW}.

\subsection{Gauge groups and matter fields}

Any 6D supergravity theory with ${\mathcal N} = (1, 0)$ supersymmetry is
characterized by the following discrete data

\begin{eqnarray*}
T & & T \in{\mathbb Z},T \geq 0\hspace*{0.1in} \makebox{[number of
    anti-self-dual tensor fields]}\\
G &  & \makebox{compact finite-dimensional Lie group [gauge group]}\\
M & & \makebox{finite-dimensional representation of $G$ [matter content]}
\end{eqnarray*}

This data describes the set of dynamical quantum fields in the
supergravity theory.  As mentioned above, the fields in the theory
live in supersymmetry multiplets--that is, in representations of the
supersymmetry algebra of the theory.  There are four types of
supersymmetry multiplets appearing in ${\mathcal N} = (1, 0)$
theories.  Each theory contains a single ``gravity'' multiplet.  The
gravity multiplet includes
among its bosonic components a symmetric tensor field $g_{\mu
\nu}$, which describes the metric on the 6D space-time manifold ${\mathcal
M}$, and an antisymmetric two-form field $B^0_{\mu \nu}$, which
satisfies a self-dual condition  $ \sim dB^0 ={}^*dB^0$.  $T$
denotes the number of ``tensor'' multiplets, which each include an
anti-self-dual two-form field $B^i_{\mu \nu}$ and a scalar field
$\phi^i, i = 1, \ldots T$.  Each model has some some number of ``vector'' multiplets,
which contain one-form gauge fields $A_\mu$ describing a connection on
a principal $G$-bundle over ${\mathcal M}$.  Each model also has a set
of ``hyper'' multiplets, each containing four scalar fields
$\varphi^a$.  These scalar fields parameterize a quaternionic K\"ahler
manifold, and transform under a (generally reducible) representation
$M$ of the gauge group $G$.  Each multiplet, in addition to the
bosonic fields just mentioned, also contains fermionic (Grassmann)
fields related to the bosonic fields through supersymmetry.  We will
not be concerned with many details of the field structure in the
analysis here; the principal data we are concerned about are the data
$T, G, M$ characterizing the number of tensor multiplets, the gauge
group, and the matter representation.  Furthermore, we will focus on
the nonabelian structure of the gauge group, taking $G$  to be
semisimple and ignoring $U(1)$ factors.  We write $G$ as a product of simple factors $H_i$
\begin{equation}
G = H_1 \times H_2 \times \cdots \times H_k \,.
\end{equation}
We denote the number of vector multiplets associated with generators
of $G$ by $V =$ dim $G$, and the number of hypermultiplets transforming
(trivially or nontrivially) under $G$ by $H =$ dim $M$.

\subsection{Anomalies}

Not all possible combinations of  discrete data $T, G, M$  can be
associated with consistent supersymmetric quantum theories of gravity.
Quantum anomalies impose strong constraints on the set of possible
theories.  In fact, the condition that a theory be free of anomalies,
along with a simple physical sign constraint on the structure of the
gauge fields, is sufficient to prove that when $T < 9$ there are only
a finite number of possible nonabelian gauge groups $G$ and matter
representations $M$ possible for a consistent quantum supergravity
theory.  This was proven for $T = 1$ in \cite{finite}, and for $T < 9$
in \cite{tensors}.  After summarizing the anomaly conditions in this
subsection, we present the key arguments in the proof of finiteness in
the following subsection.

A quantum anomaly is a breakdown at the quantum level of a
classical symmetry.
One way to understand anomalies (Fujikawa \cite{Fujikawa}) is in terms
of the path integral formulation of a quantum theory.  A quantum
theory containing fermion fields $\psi, \bar{\psi}$
can be defined through a path integral
\begin{equation}
Z = \int {\mathcal D} \psi{\mathcal D} \bar{\psi} e^{iS (\psi, \bar{\psi})},
\end{equation}
where the action $S (\psi, \bar{\psi}) = \bar{\psi}\slashed{D} \psi +
\cdots$ contains a Dirac operator $\slashed{D}$ and is invariant under
local gauge transformations $\psi \rightarrow g \psi$.  For chiral
fermionic fields, nontrivial transformation of the measure factor
${\mathcal D} \psi{\mathcal D} \bar{\psi}$ leads to a quantum violation of
the classical gauge symmetry.  In terms of Feynman diagrams, such
quantum anomalies appear in one-loop computations, and cannot be
removed by any choice of the regularization prescription used to remove
infinities from the quantum calculation.  For gravitational theories, chiral
fermions and self-dual/anti-self-dual antisymmetric tensor fields lead
to similar anomalies associated with violations of local
diffeomorphism invariance \cite{ag-Witten}.  

As was shown in
\cite{Atiyah-singer-physics, ag-Ginsparg},  gauge and
gravitational anomalies are determined by characteristic classes
associated with the index of the Dirac operator coupled to the
appropriate vector potential.  
The anomaly in $D$ dimensions is associated with an index in $D + 2$
dimensions.  Thus, in particular, gauge, gravitational, and mixed
gauge-gravitational anomalies in six dimensions are associated with an
8-form $I_8$ containing terms built from the geometric and Yang-Mills
curvatures, such as $\tr R^4,\tr F^4,\tr
R^2\tr F^2, \ldots$.  It was shown by Green and Schwarz
\cite{Green-Schwarz} that the anomaly in some 10-dimensional
supergravity theories can be cancelled by classical (tree-level) terms
associated with exchange of quanta associated with antisymmetric
two-form fields.  This mechanism was extended to six dimensions in
\cite{gsw} for theories with one tensor multiplet, and in
\cite{Sagnotti} for theories with any number of tensor multiplets.
Anomalies in six-dimensional supergravity theories with a single
supersymmetry can be cancelled by the Green-Schwarz-Sagnotti mechanism
if the anomaly polynomial $I_8$ can be written in the form
\begin{equation}
I_8(R,F) = \frac{1}{2} \Omega_{\alpha\beta} X^\alpha_4 X^\beta_4
\label{eq:factorized-anomaly}
\end{equation}
where $X^\alpha_4$ is a 4-form taking the form
\begin{equation}
X^\alpha_4 = \half a^\alpha \tr R^2 +  \sum_i b_i^\alpha \ 
\left(\frac{2}{\lambda_i} \tr F_i^2 \right) \,.
\end{equation}
Here, $a^\alpha$ and $b_i^\alpha$ are
vectors in the space $\R^{1,T}$ carrying a symmetric bilinear form
$\Omega_{\alpha\beta}$.  The normalization factor $\lambda_i$ is
chosen depending on the gauge factor $H_i$ such that minimal
instantons have compatible normalization factors.  For example,
$\lambda_{SU(N)} = 1,$ while $\lambda_{E_8} = 60$.

The anomaly polynomial does not specify the vectors $a^\alpha,
\ b_i^\beta$, but only constrains the $SO(1,T)$ invariant quantities
$a^2, a \cdot b_i, b_i \cdot b_j$, where the inner product is taken
with respect to the form $\Omega$.  
A detailed calculation of the anomaly arising from all chiral fields
in the theory (see  \cite{gsw, Erler, Schwarz} for details) gives
the form of the complete anomaly polynomial as
a function $I (R, F)$ of the gravitational and Yang-Mills curvature
two-forms $R$ and $F$.

In order for the anomaly to
factorize as in (\ref{eq:factorized-anomaly}), $I_8$ cannot contain
terms proportional to $\tr R^4$,  which gives the condition
\begin{equation}
H-V = 
273- 29T \,.
\label{eq:bound}
\end{equation} 
The $\tr F^4$ contribution to the total anomaly must also cancel for
each gauge group factor,
giving the condition
\begin{equation}
 B^{i}_{Adj}  =  \sum_R x^i_{R} B^i_{R} \label{eq:f4-condition}
\end{equation}
where we use the group theory coefficients $A_R,B_R,C_R$  defined through
\begin{align}
\tr_R F^2 & = A_R  \tr F^2 \\
\tr_R F^4 & = B_R \tr F^4+C_R (\tr F^2)^2
\end{align}
and where $x_R^i$ denotes the number of hypermultiplets transforming
in irreducible representation $R$ under gauge group factor $H_i$.

The remaining conditions that the anomaly factorize relate inner
products between the vectors $a, b_i$ to group theory coefficients and
the representations of matter fields
\begin{align}
a \cdot a & =  9 - T  \label{eq:aa}\\ 
a \cdot b_i & =  \frac{1}{6} \lambda_i  \left( A^i_{adj} - \sum_R x^i_R A^i_R\right)\label{eq:ab-condition} \\
b_i\cdot b_i & = -\frac{1}{3} \lambda_i^2 \left( C^i_{adj} - \sum_R x^i_R C^i_R  \right)  \label{eq:bii-condition}\\
b_i \cdot b_j & =  \lambda_i \lambda_j \sum_{RS} x_{RS}^{ij} A_R^i
A_S^j
\label{eq:bij-condition}
\end{align}
where $x_{RS}^{ij}$ 
denotes the number of hypermultiplets transforming in irreducible
representations $R, S$ under the gauge group factors $H_i, H_j$.

\section{Constraints from anomalies}
\label{sec:constraints}

The conditions (\ref{eq:bound}-\ref{eq:bij-condition}) place strong
constraints on the range of possible theories compatible with anomaly
cancellation, particularly for small values of $T$.  For one thing,
these conditions imply that the inner products $a \cdot a, a \cdot
b_i, b_i \cdot b_j$ are all integral and hence that the vectors $a,
b_i \in \R^{(1, T)}$ span an integral lattice $\Lambda$ (which may be
degenerate, for example if several $b_i$ are equal).  For gauge group
factors $H_i$ with an irreducible quartic invariant (nonzero $B$'s),
the integrality of the inner products follows from group theory
identities which show that, {\it e.g.}, $3 | \lambda_i^2 C^i_R$ for
all representations $R$.  For group factors with no irreducible
quartic invariant, such as $SU(2), SU(3), E_8$, integrality of the
inner products depends upon cancellation of additional global
anomalies.  Details of the proof of integrality are given in
\cite{tensors}.

Beyond the integral lattice structure imposed by
(\ref{eq:aa}-\ref{eq:bij-condition}), these conditions also
provide strong bounds on the set of possible gauge groups and matter
content.  To give a finite bound to the set of allowed $G, M$ for
small $T$ we need to impose an additional condition, associated with
the constraint that each gauge group component has a sensible physical
description without instabilities.  This condition amounts to the
constraint that there exists a vector $j \in\R^{1, T}$ such that
\begin{equation}
j \cdot b_i > 0 \; \;\forall i,
\;\;\;\;\; j^2 = 1 \,.
\label{eq:jb}
\end{equation}
The vector $j$ in this equation corresponds to the scalar fields in
the $T$ tensor multiplets of the supergravity theory.

Given these conditions, we can prove
\begin{theorem}
\label{theorem:finite}
For $T < 9$, there are a finite number of distinct  nonabelian
gauge groups $G$
and matter representations $M$ satisfying
(\ref{eq:bound}-\ref{eq:bij-condition}), such that
(\ref{eq:jb}) holds for some $j$.
\end{theorem}

This theorem is proven in \cite{finite} for $T = 1$, and in
\cite{tensors} for $T < 9$.  The proof essentially relies upon
geometry in the $(1, T)$ signature space containing the vectors $a,
b_i, j$.  We present here the main arguments in the proof.
\begin{proof}

The proof proceeds by contradiction.  We assume that there is an
infinite family of models with
nonabelian gauge groups $\{G_\gamma\}$.  For any given model in the
family we decompose the gauge group into a product of simple factors
$G_\gamma= H_1 \times H_2 \times \cdots \times H_k $.  There are a
finite number of groups $G$ with dimension below any fixed bound.  For
each fixed $G$, there are a finite number of representations whose
dimension is below the bound (\ref{eq:bound}) on the number of
hypermultiplets.  Thus, any infinite family $\{G_\gamma\}$ must
include gauge groups of arbitrarily large dimension.  We divide the
possibilities into two cases.

\begin{enumerate}
\item The dimension of the simple factors in the groups $\G_\gamma$ is
  bounded across all $\gamma$.  In this case, the number of simple
  factors is unbounded over the family.

\item The dimension of at least one simple factor in $\G_\gamma$ is
unbounded.  For example, the gauge group is of the form
$\G_\gamma=SU(N_\gamma)\times \tilde{G}_\gamma$, where $N_\gamma
\rightarrow \infty$.
\end{enumerate}

{\bf Case 1}: In this case we can rule out infinite families with
bounded values of $T$.  The dimension of each factor $H_i$ is bounded,
say by dim $H_i \leq D$.  Assume that we have an infinite sequence of
models whose gauge groups have $ N_\gamma$ factors, with $N_\gamma$
unbounded.  Consider one model in this infinite sequence, with $N$
factors.  We divide the factors $H_i$ into 3 classes, depending upon
the sign of $b_i^2$:

\begin{enumerate}
\item {\bf Type Z}: $b_i^2 = 0$
\item {\bf Type N}: $b_i^2 < 0$
\item {\bf Type P}: $b_i^2 > 0$
\end{enumerate}

Since the dimension of each factor is bounded,
the contribution to $\nhv$ from $-V$ is bounded below by $-N D$.
For fixed $T$ the total number of hypermultiplets is then bounded
by 
\begin{equation}
H \leq 273-29T + N D \equiv B \sim{\mathcal O} (N)\,.
\end{equation}
This means that the dimension of any irreducible component of $M$ is
bounded by the same value $B$.  The number of gauge group factors
$\lambda$ under which any matter field can transform nontrivially is
then bounded by $2^\lambda \leq B$, so $\lambda \leq{\mathcal O} (\ln
N)$.

Now, consider the different types of factors.  Denote the number of
type N, Z, P factors by $N_{N, Z, P}$, where
\begin{equation}
N = N_N + N_Z + N_P \,.
\end{equation}
We can write the
$b_i$'s in a (not necessarily integral)
basis where the inner product matrix takes the form
$\Omega = {\rm diag} (+ 1, -1, -1, \ldots)$ as
\begin{equation}
b_i = (x_i, \vec{y}_i) \,.
\label{eq:bxy}
\end{equation}
For any type P factor, $|x_i | > | \vec{y}_i |$, so $b_i \cdot b_j >
0$ for any pair of type P factors.  Thus, there are hypermultiplets
charged under both gauge groups for every pair of type P factors.  A
hypermultiplet charged under $\lambda \geq 2$ gauge group factors
appears in $\lambda (\lambda -1)$ (ordered) pairs $(i, j)$ with $b_i
\cdot b_j > 0$, and contributes at
least $2^\lambda$ to the total number of hypermultiplets $H$.  Each
ordered pair under which this hypermultiplet is charged then
contributes at least
\begin{equation}
\frac{2^\lambda}{ \lambda (\lambda -1)} \geq 1
\end{equation}
to the total number of hypermultiplets $H$.  It follows that the $N_P
(N_P -1)$ pairs under which at least one hypermultiplet is charged
contribute at least $N_P (N_P -1)$ to $H$, so
\begin{equation}
N_P (N_P -1) \leq B
\end{equation}
Thus,
\begin{equation}
N_P <\sqrt{B} + 1 \leq{\mathcal O} (\sqrt{N})
 \label{eq:order-np}
\end{equation}
which is much smaller than $N$ for large $N$.  
So most of the $b_i's$ associated with gauge group factors in any
infinite family must be type $Z$ or type $N$.

Now consider type N
factors.  
Any set of $r$ mutually orthogonal type $N$ vectors defines an
$r$-dimensional negative-definite subspace of $\R^{1, T}$.
This means, in particular, that we cannot have $T + 1$ mutually
orthogonal type $N$ vectors.  
If we have $N_N$ type $N$ vectors, we can define a graph whose nodes
are the type $N$ vectors, where an edge connects every two nodes
associated with perpendicular vectors.  Tur\'an's theorem \cite{Turan}
states that the maximum number of edges on any graph with $n$ vertices
which does not contain a subset of $T + 1$ completely connected
vertices is
\begin{equation}
(1-1/T) n^2/2
\end{equation}
where the total number of possible edges is $n (n -1)/2$.  Thus,
applying this theorem to the graph described above on nodes associated
with type $N$ vectors, the number of ordered pairs with charged
hypermultiplets must be at least
\begin{equation}
\frac{N_N^2}{T} -N_N   \,.
\end{equation}
It then follows that, assuming $T$ is fixed,
\begin{equation}
N_N \leq \sqrt{TB} + T  \sim{\mathcal O} (\sqrt{N})\,.
 \label{eq:order-nn}
\end{equation}

Finally, consider type Z factors.  Vectors $b_i, b_j$ of the form
(\ref{eq:bxy}) associated with two type Z factors 
each have $| x_i | = | \vec{y}_i |$ and
have a
positive inner product  unless they are parallel, in which case $b_i
\cdot b_j  = 0$. Denote by $\mu$ the size of the largest collection of
parallel type Z vectors.  Each type Z vector is perpendicular to fewer
than $\mu$ other type Z vectors, so there are at least $N_Z
(N_Z-\mu)$ ordered
pairs of type Z factors under which there are charged hypers. We must then
have
\begin{equation}
N_Z (N_Z-\mu) = (N_Z-\mu) (N -N_P-N_N) \leq  B \,.
\end{equation}
But from (\ref{eq:order-np}, \ref{eq:order-nn}) this means that
$N_Z-\mu$ is of order at most ${\mathcal O} (1)$ (and is bounded by
$D$ as $N \rightarrow \infty$), while $N_Z$ is of order ${\mathcal O}
(N)$.  Thus, all but a fraction of order $1/N$ of the type Z factors
have vectors in a common parallel direction.  In \cite{finite}, we
carried out a case-by-case analysis demonstrating that all group +
matter configurations which give type Z factors have a positive value
for $\nhv$.  All of the $\mu$ factors associated with the largest
collection of parallel type Z vectors thus contribute positively to
$\nhv$, counting matter charged under these group factors only once.
The total contribution to $\nhv$ is then bounded by
\begin{equation}
\nhv  > \mu - \left[(N_Z-\mu) + N_P + N_N \right] (D)
\sim {\mathcal O} (N)
\end{equation}
which exceeds the bound $H -V \leq 273-29T$ for sufficiently large
$N$.  Thus, we have ruled out case 1 by contradiction for any fixed $T
> 0$, and in consequence ruled out any infinite family of the type of
case 1 with bounded values of $T$.

\vspace*{0.1in}

{\bf Case 2}:

We now consider the possibility of infinite families with
factors of unbounded size.
We assume that we have an infinite family of models with gauge groups
of the form
$\G_\gamma=H_\gamma \times \tilde{G}_\gamma$ where dim $H_\gamma
\rightarrow \infty$.
For $SU(N)$ the $F^4$ anomaly
cancellation condition
\begin{eqnarray}
B_{\rm Adj} =2N = \sum_R x_R B_R \label{eq:SU-trace}
\end{eqnarray}
can only be satisfied at large $N$ when the number of multiplets $x_R$
vanishes in all representations other than the fundamental, adjoint,
and two-index antisymmetric and symmetric representations (we consider
representations and their conjugates to be equivalent for the purposes
of this analysis).  
This follows from the fact that for all other representations, $B_R$
grows faster than $N$.
For the
representations just listed, indexed in that order, (\ref{eq:SU-trace}) becomes
\begin{eqnarray}
2N = x_1+2N x_2+(N-8)x_3+(N+8)x_4 \,. \label{eq:SU-trace-1}
\end{eqnarray} 
The solutions to this equation at large $N$, along with the
corresponding solutions for the other classical groups $SO(N), Sp(N)$
are easy to tabulate, and are
listed in Table~\ref{table:solutions}.  We discard solutions
$(x_1, x_2, x_3, x_4) = (0, 1, 0, 0)$ and $(0, 0, 1, 1)$, where $a
\cdot b_i = b_i^2 = 0$ since for $T < 9$ these relations combined with
$a^2 > 0$ imply that
$b_i = 0$ and therefore that the kinetic term for the gauge field is
identically zero.
\begin{table}
\centering
	\begin{tabular}{|c|c|c|c|c|}
	\hline
	Group & Matter content & $\nhv$ & $a \cdot b$ & $b^2$ \\
	\hline
	\multirow{4}{*}{$SU(N)$} & $2N\ {\tiny\yng(1)}$ & $N^2+1$& 0
	& -2\\
	& $(N+8)\ {\tiny\yng(1)}+1\ {\tiny\yng(1,1)}$ & $
	\frac{1}{2}N^2+\frac{15}{2}N+1 $& 1 & -1\\
	& $(N-8)\ {\tiny\yng(1)}+1\ {\tiny\yng(2)}$ & $
	\frac{1}{2}N^2-\frac{15}{2}N+1 $& -1 & -1\\
	& $16\ {\tiny\yng(1)}+2\ {\tiny\yng(1,1)}$ & $ 15N+1 $ & 2 & 0\\
	\hline
	$SO(N)$ & $(N-8)\ {\tiny\yng(1)}$ &
	$\frac{1}{2}N^2-\frac{7}{2}N$ & -1 & -1 \\
\hline
\multirow{2}{*}{$Sp(N/2)$} & $(N+8)\ {\tiny\yng(1)}$ &
	$\frac{1}{2}N^2+\frac{7}{2}N$ & 1 & -1 \\
	 & $16\ {\tiny\yng(1)}+1\ {\tiny\yng(1,1)}$ & $15N-1$ & 2 & 0
	 \\ \hline
\end{tabular} \caption{Allowed charged matter for
	 an infinite family of models with gauge group $H(N)$.  The
	 last two columns give the values of $ a \cdot b, b^2$ in the
 anomaly polynomial $I_8$.}  \label{table:solutions}
\end{table}
The contribution to $\nhv$ from each
of the group and matter combinations in Table~\ref{table:solutions}
diverges as $N \rightarrow \infty$.  This cannot be cancelled by
contributions to $-V$ from an infinite number of factors, for the
same reasons which rule out case 1.  Thus, any infinite family must
have an infinite sub-family, with gauge group of the form
$\hat{H}(M)\times H(N) \times \tilde{G}_{M, N}$, with both $M,
N\rightarrow \infty$.  For any factors $H_i, H_j$ with $a \cdot b_i, a
\cdot b_j \neq 0$, in a (non-integral) basis where $\Omega = {\rm
diag} (+ 1, -1, -1, \ldots)$, and $a = (\sqrt{a^2}, 0, 0, \ldots)$
writing
\begin{equation}
b_i = (x_i, \vec{y}_i) 
\label{eq:bxy-2}
\end{equation}
with $x_i = a \cdot b_i/\sqrt{a^2}$ we have 
\begin{equation}
x_ix_j = (a \cdot b_i) (a \cdot b_j)/a^2\geq b_i \cdot b_j  = \sum_{R, S}x_{R S} A_R A_S \,.
\end{equation}
Since $x_ix_j$ is constant for  any infinite family of pairs
$\hat{H} (M), H (N)$, while
$A_R$ grows for all representations besides the fundamental, the only
possible fields charged under more than one of the infinite factors in
Table~\ref{table:solutions} are bifundamentals.

We now consider all possible infinite families built from products of
groups and representations in Table~\ref{table:solutions} which have
bifundamental fields and bounded $\nhv$.  There are 5 such
combinations with two factors.  These combinations were enumerated in
\cite{finite}, and are listed in Table 4 in that paper.  These
combinations include two infinite families shown to satisfy anomaly
factorization by Schwarz \cite{Schwarz}, as well as three other
similar families.  For example, the simplest such family consists of
an infinite sequence of models with gauge group $SU(N) \times SU(N)$
and two matter fields transforming in the bifundamental ($N, N$)
representation.  This model has $H = 2 N^2, V = 2 (N^2 -1)$ so $\nhv =
2$ for each model in the infinite family.  In addition to the infinite
families built from products of two factors with unbounded dimension,
when $T > 1$ there are several infinite families arising from products
of three factors with unbounded dimension.  For example, there is such
a family of models with gauge group $G = SU(N -8) \times SU(N) \times
SU(N+ 8)$.  The other possibilities are described in \cite{tensors}.
For $T < 9$, the models in all of these infinite families with
unbounded factors are unacceptable because the gauge kinetic term for
one of the factors is always unphysical.  This can be shown as
follows: For each two-factor infinite family we have two vectors $b_1,
b_2$ which satisfy $a \cdot (b_1 + b_2) = 0$ and $(b_1 + b_2)^2 = 0$.
But when $a^2 > 0$ these conditions imply $b_1 + b_2 = 0$, so that $j
\cdot b_1$ and $j \cdot b_2$ cannot both be positive.  For example,
for the theory found by Schwarz with gauge group $SU(N) \times SU(N)$
with two bifundamental fields, we have $a\cdot b_1 = a \cdot b_2 = 0,
b_1^2 = b_2^2 = -2, b_1 \cdot b_2 = 2$, from which it follows that
$b_1 = -b_2$.  Essentially the same argument works for each
three-factor infinite family, where $a \cdot (b_1 + b_2 + b_3) = 0$.
This method of proof breaks down when $T > 8$, where $a^2 \leq 0$,
since then $a \cdot b = b^2 = 0$ is not sufficient to prove $b = 0$.
In the following section we give an example of an infinite family of
models with no clear inconsistency at $T = 9$.

This proves case 2 of the analysis.  So we have proven that for $T< 9$
there are a finite number of distinct gauge groups and matter content
which satisfy anomaly cancellation with physical kinetic terms for all
gauge field factors.  We have ruled out infinite families with
unbounded numbers of gauge group factors at any finite $T$, though
infinite families with unbounded numbers of factors
satisfying the anomaly and group theory sign constraints 
can exist when $T$ is
unbounded, as discussed in \cite{tensors}.  We have not ruled out
infinite families with a finite number of gauge group factors which
become unbounded at finite $T > 8$.  Indeed, we give an explicit
construction of such a family in Section \ref{sec:examples}.
\vspace*{0.1in}

\end{proof}

Theorem (\ref{theorem:finite}) strongly constrains the range of
possible supergravity theories, at least when $T < 9$.  Furthermore,
the analysis in the proof of the theorem can be used to give explicit
algorithms for systematically enumerating all the possible gauge
groups and matter content which can appear in acceptable theories, as
discussed in \cite{finite, KMT}.
 
\section{6D supergravities from F-theory}
\label{sec:F-theory}

\subsection{F-theory models from elliptically fibered Calabi-Yau
  threefolds}

There are many ways in which string theory can be used to construct
supergravity theories in various dimensions.  In general, such
constructions involve starting with a 10-dimensional string theory,
and ``compactifying'' the theory by wrapping $10 - D$ dimensions of
the space-time on a compact $(10-D)$-dimensional manifold to give a theory
in $D$ dimensions which behaves as a supergravity theory at low
energies.  At this point in time, we do not have any fundamental
``background-independent'' definition of string theory, in which all
the different compactifications of the theory arise on equal footing.
Rather, string theory consists of an assemblage of tools, including
low-energy supergravity, perturbative string theory, and
nonperturbative structures such as D-branes and duality symmetries,
which give an apparently consistent way of constructing and relating
the various approaches to string compactification.

One of the most general approaches to string compactification is
``F-theory'' \cite{F-theory, Morrison-Vafa}.  Technically, F-theory
describes certain limits of string compactifications in which certain
information (such as K\"ahler moduli) is lost.  Mathematically, an
F-theory compactification to $D$ dimensions is characterized by an
elliptically fibered Calabi-Yau manifold of dimension $12 -D$.  The
data needed to define an F-theory compactification to six dimensions
consist of a complex surface $B$ which acts as the base of the
fibration, and an elliptic fibration with section over $B$.  The
elliptic fibration may have singularities, as long as such
singularities can be resolved to give a total space which is
Calabi-Yau.  The structure of such an elliptic fibration is generally
described by a Weierstrass model
\begin{equation}
y^2 = x^3 + f(u, v) xz^4 + g(u,v)z^6 \label{eq:weierstrass}
\end{equation}
where $u, v$ are coordinates on the base $B$.
The functions $f,g$
are sections of the line bundles $-4K$ and $-6K$ respectively, where
$K$ is the canonical bundle of $B$.

The gauge group and matter structure of the low-energy theory
associated with an F-theory compactification on an elliptically
fibered Calabi-Yau threefold are determined by the singularity
structure of the fibration.  Simple nonabelian gauge group factors
$H_i$ in the associated 6D theory are associated with codimension one
loci in the base where the fiber degenerates.  Such codimension one
singularities were classified by Kodaira \cite{Kodaira}.  These
singularities can be associated with A-D-E Dynkin diagrams and give
rise to the corresponding simply-laced gauge group factors.  Nontrivial
monodromies about codimension 2 singularities expand the possible
gauge group factors to include the non-simply laced groups, $Sp(N),
F_4$ and $G_2$ \cite{Bershadsky-all}. Given a Weierstrass form for the
elliptic fibration, the singularity types can be determined in terms of the orders of vanishing of $f, g,$
and the discriminant locus
\begin{equation}
\Delta = 4f^3 + 27g^2,
\end{equation}
which is a section of $-12K$.  The resulting gauge group can be then
determined via Tate's algorithm \cite{Tate}.  For example, a
singularity of Kodaira type $I_n$, corresponding to the Dynkin diagram
$A_{n -1}$, arises on a codimension one singularity locus where
neither $f$ nor $g$ vanishes, but $\Delta$ vanishes to order $n$.
Such a singularity gives a gauge group factor $SU(n)$ when the base of
the fibration is (complex) one-dimensional, and $SU(n)$ or $Sp(n/2)$
in the low-energy 6D theory when the base is 2 dimensional.  A
complete list of the possible singularity types and associated orders
of vanishing of $f, g, \Delta$ can be found in, for
example,\cite{Morrison-Vafa, Bershadsky-all}.  Each nonabelian gauge
group factor $H_i$ is associated with a singularity locus on an
effective irreducible divisor $\xi_i \in H_2 (B;\Z)$.

Matter fields (hypermultiplets) in the low-energy 6D theory are
associated with codimension two singularities in the elliptic
fibration.  Generally such singularities arise at intersections of
divisors associated with codimension one singularities.  For example,
an $A_{n -1}$ singularity intersecting an $A_{m -1}$ singularity gives
rise to a codimension two $A_{n + m -1}$ singularity, where $\Delta$
vanishes to order $n + m$.  The resulting matter fields transform in
the fundamental of the associated $SU(n)$ gauge group factor and the
(anti)-fundamental of the associated $SU(m)$ gauge group factor.
(This can be thought of as splitting the adjoint representation of
$SU(n + m)$ as a representation of $SU(n) \times SU(m)$).  The
singularities associated with possible matter representations which
can arise in this fashion have not been fully classified, but an
analysis of many such representations appears in
\cite{Grassi-Morrison, Grassi-Morrison-2}.

As discussed in \cite{Morrison-Vafa}, it was shown by Kodaira for a
base of complex dimension 1, and subsequently generalized to higher
dimension by other authors, that the total space of the elliptic
fibration can be resolved to a Calabi-Yau manifold when $-12K =
\Delta$, where $\Delta$ is the total divisor class of the singular
(discriminant) locus.  The component of the discriminant locus for
each nonabelian factor is given by an effective irreducible divisor
$\xi_i$ with a multiplicity $\nu_i$.  For example, for an $A_{n -1}$
singularity, $\nu_i = n$.  The residual singularity locus, which is
not associated with nonabelian gauge group structure, is associated
with another effective divisor $Y$, so that the Kodaira relation is
\begin{equation}
-12K = \Delta  = \sum_{i}\nu_i \xi_i + Y\,.
\end{equation}

In this fashion, a wide range of 6D supergravity theories can be
constructed from particular choices of elliptically fibered Calabi-Yau
threefolds.  In all such situations, the resulting gauge group and
hypermultiplet matter content automatically satisfy the anomaly
cancellation conditions \cite{Sadov, Grassi-Morrison}.  We now discuss
the connection between the 
data of the low-energy theory and that of
the F-theory compactification.

\subsection{Mapping 6D supergravities to F-theory}

To identify the set of low-energy 6D supergravity models which are
compatible with F-theory, we would like to have a systematic approach
to constructing an F-theory model which realizes a particular
low-energy gauge group $G$ and matter representation $M$.  In fact, as
shown in \cite{Sadov, Grassi-Morrison}, the anomaly cancellation
conditions give a close correspondence between the data of a geometric
F-theory construction and the corresponding 6D supergravity theory.
This correspondence characterizes a map from low-energy data
to topological F-theory data \cite{KMT, tensors}; identifying those
cases where such a map
cannot be consistently defined allows us to determine which $G, M$
cannot be found through any F-theory construction.

The connection between the 6D supergravity theory and F-theory
data is given by the following correspondences
\cite{Sadov, Grassi-Morrison, KMT, tensors}: the number of tensor
multiplets $T$ in the 6D theory is equal to $h^{1, 1}-1$ of the base
$B$ of the F-theory elliptic fibration.  The lattice $\Lambda$
determined by the anomaly cancellation conditions through
(\ref{eq:ab-condition}-\ref{eq:bij-condition}) can be embedded into
the integral second homology of $B$
\begin{eqnarray}
\Lambda & \hookrightarrow & H_2 (B;\Z) 
\label{eq:map}
\end{eqnarray}
in such a way that
\begin{eqnarray}
a & \rightarrow &  K \label{eq:a-map}\\
b_i & \rightarrow &  \xi_i\label{eq:b-map}
\end{eqnarray}
Thus, for example, we have
\begin{equation}
a \cdot a =9-T = 10-h^{1, 1} = K^2 \,.
\end{equation}

For an elliptic fibration which resolves to a smooth Calabi-Yau
threefold, the possible smooth bases $B$ consist of $\P^2$
and the Hirzebruch surfaces $\F_m$ ($m \leq 12$) and  blow-ups of
these bases, as well as the
somewhat trivial cases of K3 and the Enriques surface \cite{Morrison-Vafa}.
For example, at $T = 1$, the set of possible F-theory bases are $\F_m,
m \leq 12$.  In this case, a basis for $H_2 (B;\Z)$ is given by $D_v,
D_s$ where $D_v^2 = -m, D_v \cdot D_s = 1, D_s^2 = 0$.  As shown in
\cite{KMT}, in this case, the vector $b$ associated with a given gauge
group factor maps to the divisor
\begin{equation}
b \; \rightarrow
\;\xi =\frac{\alpha}{ 2}  (D_v + \frac{m}{2} D_s)
+ \frac{ \tilde{\alpha}}{ 2}  D_s \,,
\label{eq:map-1}
\end{equation}
where $\alpha, \tilde{\alpha}$ satisfy $-a \cdot b = \alpha +
\tilde{\alpha}, b^2 = \alpha \tilde{\alpha}/2$.

In the following section we give some explicit examples of 6D
supergravity theories and the corresponding F-theory constructions.
In some cases, there is no map of the form (\ref{eq:map}) where $a,
b_i$ map to effective divisors in any $B$ with the correct properties for
an F-theory compactification.
For these models, which are apparently consistent from the
low-energy.of view and
yet have no F-theory construction, we can identify which aspect of the
F-theory structure breaks down.  We include some examples of this type
in the following section.  In some other cases, there are multiple possible
F-theory realizations of the specified 6D gauge group $G$ and matter
representation $M$.  Thus, the correspondence does not give a uniquely
defined map from low-energy data to F-theory.  It does, however, give
us a framework for analyzing which low-energy theories admit an
F-theory realization.
(Note that in those cases where the gauge group and matter content are
not sufficient to uniquely determine the F-theory construction,
further information about the low-energy theory may single out a
particular F-theory model.)

\section{Examples}
\label{sec:examples}

In this section we consider some examples of possible gauge groups and
matter representations for 6D supergravities.  We begin with two
examples of theories which seem to admit consistent embeddings in
F-theory, and then consider two specific examples and one infinite
family of theories which cannot be realized using standard F-theory
techniques (or any other known string construction).
\vspace*{0.05in}

\noindent
{\bf Example 1:} $T = 0, G =SU(N),  M = 3 \times {\tiny\yng(1,1)}
+(24 -N) \times {\tiny\yng(1)}$
\vspace*{0.05in}

A simple class of examples involve theories with no tensor multiplets
($T = 0$).  In this case, $a^2 = 9$, and the space $\R^{1, T} =\R^1$ is just
one-dimensional Euclidean space, so $-a= (3)$ in an integral basis.
Any $SU(N)$ factor is associated with a vector $b = (k)$.  For
simplicity, we assume that the only matter consists of $F$
hypermultiplets transforming in the fundamental representation and $A$
hypermultiplets in the two-index antisymmetric representation
$({\tiny\yng(1,1)})$.  We can use the group theory coefficients
\begin{equation}
\begin{array}{lll}
   A_f = 1,  
& B_f = 1, 
& C_f = 0  \\
    A_a = N-2,  \hspace*{0.1in}
& B_a = N -8, \hspace*{0.1in}
& C_a = 3\\
 A_{\rm Adj} =2 N, 
& B_{\rm Adj} = 2 N,
& C_{\rm Adj} =
6
\end{array} \label{eq:coefficients}
\end{equation}
for $SU(N)$
to write the anomaly cancellation equations:
\begin{eqnarray}
0 & = &2 N - F - A (N -8) \nonumber\\
-a \cdot b =  3k & = &  -(2 N -F-A (N -2))/6 = A\\
b^2 =k^2 & = & (-6 + 3A)/3 = A-2 \nonumber
\end{eqnarray}
For $k = 1$, we have  solutions for $A = 3$, with $F = 24 -N$ for $N
\leq 24$.  This gives a set of models with simple gauge group $G =
SU(N)$ which satisfies all anomaly cancellation conditions.  The gauge
kinetic term condition is satisfied for any positive $j = (j_0)$.

For the corresponding F-theory construction, we have $\P^2$ as the
only possible base with $h^{1, 1} = T + 1 = 1$.
The second homology of $\P^2$ is generated by the cycle $H$,
corresponding to the hyperplane divisor, satisfying $H \cdot H = 1$.
The map from $\Lambda  \hookrightarrow H_2 (B;\Z) $ is then given by
the
trivial map $(1) \rightarrow H$, so that 
\begin{eqnarray}
-a = (3) & \rightarrow & 3H = -K \\
b = (1) & \rightarrow &  H \,.\nonumber
\end{eqnarray}
This gives the topological data associated with the corresponding
F-theory construction.  To verify that there exists an F-theory model
from an elliptic fibration associated with this data one must
construct a Weierstrass model with singularity type $A_{N -1}$ on the
divisor $H$.  We discuss explicit construction of Weierstrass models
in the following section.  One can similarly construct a variety of
anomaly-free models with gauge group of the form $G =
\prod_{i}SU(N_i)$ and matter in fundamental, antisymmetric, and
bifundamental representations.  We explicitly constructed all such
models for $T = 1$ in \cite{KMT} and found that all appear to correspond to
acceptable topological data for F-theory constructions.
\vspace*{0.1in}

\noindent
{\bf Example 2:} $T = 1, G = E_8 \times E_7$
\vspace*{0.05in}

This gauge group has two simple factors, and thus two vectors $b_8,
b_7$ in $\R^{1, 1}$.  For $E_8$ and $E_7$ we have the normalization
factors $\lambda_8 = 60, \lambda_7 = 12$.  Neither $E_8$ or $E_7$ has
a fourth order invariant, so (\ref{eq:f4-condition}) is automatic.
Assume we have $F$ matter fields in the fundamental representation of
$E_7$, with no matter transforming under the $E_8$ (the fundamental
representation for $E_8$ is equivalent to the adjoint).  The $E_8$
anomaly equations state that
\begin{equation}
-a \cdot b_8 = -10, \;\;\;\;\;
b_8^2 = -12
\end{equation}
where $a^2 = 8$.  In a basis where
\begin{equation}
\Omega_{\alpha\beta} = \begin{pmatrix}
0 & 1 \\
1 & 0
\end{pmatrix} 
\label{eq:canonical-omega}
\end{equation}
we can choose $-a = (2, 2)$, and $b_8 = (1, -6)$.  The anomaly
equations for $E_7$, including the $\tr F_7^2 \tr F_8^2$ 
condition, state that
\begin{equation}
-a \cdot b_7 = 2F-6, \;\;\;\;\;
b_7^2 = 2F-8, \;\;\;\;\;
b_7 \cdot b_8 = 0 \,.
\end{equation}
These equations have the unique solution
$b_7 = (1, 6)$, realized when $F = 10$.  Thus, the anomaly
cancellation conditions for this class of theory are only realized for
this value of $F$.  In this case we can read the corresponding
F-theory model directly from (\ref{eq:map-1}):
\begin{eqnarray}
b_8 = \frac{1}{2}(\alpha_8, \tilde{\alpha}_8) & \rightarrow &  
\xi_8 =D_v + (m/2-6)D_s\\
b_7 = \frac{1}{2}(\alpha_7, \tilde{\alpha}_7) &  \rightarrow& 
\xi_7 =D_v + (m/2 + 6)D_s
\end{eqnarray}
which gives effective irreducible divisors precisely when $m = 12$.
In this case we have an acceptable F-theory construction on the base
manifold $\F_{12}$ \cite{Morrison-Vafa}, with $\xi_8 = D_v,   \xi_7 =
D_u = D_v + 12D_s$.  This model was identified in
the corresponding heterotic string formulation by Seiberg and Witten
in \cite{Seiberg-Witten}.
\vspace*{0.1in}

We now present several models which seem consistent from the point of
view of anomalies and gauge kinetic terms, but which have no known
string theory realization.
\vspace*{0.1in}

\noindent
{\bf Example 3:} $T = 1, G = SU(4), M = {\rm Adj} +
10 \times  {\raisebox{-0.1cm}[0.25cm][0.25cm]{\tiny\yng(1,1)}}+ 40 \times
\makebox{\tiny\yng(1)}$
\vspace*{0.05in}

For this model the anomaly cancellation conditions give
\begin{equation}
\Lambda =
\left(\begin{array}{cc}
a^2 &  -a \cdot b\\
-a \cdot b & b^2
\end{array} \right)
=
\left(\begin{array}{cc}
8 & 10\\
10 & 10
\end{array} \right)
\end{equation}

This lattice cannot be embedded in any unimodular lattice
\cite{tensors}.  Since the intersection form on $H_2 (B;\Z)$ must give
a unimodular lattice by  Poincar\'{e} duality, this model cannot be
embedded in F-theory.
\vspace*{0.1in}

\noindent
{\bf Example 4:} $T =  1, G = SU(8), M = {\tiny\yng(2)}$
\vspace*{0.05in}

Using the coefficients $A = 10, B = 16, C = 3$ for the symmetric
representation, we find that the anomaly conditions give
\begin{equation}
\Lambda =
\left(\begin{array}{cc}
a^2 &  -a \cdot b\\
-a \cdot b & b^2
\end{array} \right)
=
\left(\begin{array}{cc}
8 & -1\\
-1 & -1
\end{array} \right) \,.
\end{equation}

While this lattice can be embedded in $H_2 (B;\Z)$ for some F-theory
bases, the resulting divisor $b$ is not an effective irreducible
divisor.  For example, we could choose $-a \rightarrow -K = 2D_v + 3D_s$ on
$\F_1$  giving $b \rightarrow -D_v$.  But since the divisors $\xi_i$
carrying the singularity locus on the F-theory base must be
irreducible effective divisors, this will not work for constructing an
F-theory realization of this model using existing methods.
\vspace*{0.1in}

\noindent
{\bf Example 5:} $T = 9, G = SU(N) \times SU(N),
M = 2 \times ({\tiny\yng(1)},  {\tiny\yng(1)})$

For this gauge group and matter
representations, at $T = 9$ we need vectors $-a, b_1, b_2$ with inner
product matrix
\begin{equation}
\Lambda = \left(\begin{array}{ccc}
 0 & 0 & 0\\0 & -2 & 2\\0 & 2 & -2
\end{array} \right) \,.
\end{equation}
In a basis with $\Omega = {\rm
diag} (+ 1, -1, -1, \ldots)$, this can be realized through the vectors
\begin{eqnarray*}
-a & = &  (3, -1, -1, -1, -1, -1, -1, -1, -1, -1) \nonumber\\
b_1 & = & (1, -1, -1, -1, 0, 0, 0, 0, 0, 0)\label{eq:infinite-9}\\
b_2 & = & (2, 0, 0, 0, -1, -1, -1, -1, -1, -1) \nonumber
\end{eqnarray*}
This satisfies the correct gauge kinetic term
sign conditions $j \cdot b_i > 0$ for $j = (1, 0, 0, \ldots)$.
Thus, this infinite family of models cannot be ruled out by the
low-energy constraints we have imposed here.

These models are not valid in F-theory, however, when $N > 12$, since
in that case
\begin{equation}
Y = -12K-N (\xi_1 + \xi_2)= (N-12) K
\end{equation}
which cannot be effective for $N > 12$ since $-K$ is effective.

\section{String universality?}
\label{sec:conclusions}

The results presented so far demonstrate that, although anomaly
cancellation and other known constraints provide substantial limits on
the range of gauge groups and matter representations which can be
realized in supersymmetric six-dimensional quantum theories of
gravity, not all models which satisfy these constraints can at this
time be realized in string theory.  Some of the examples presented in
the previous section seem like counterexamples to the string
universality conjecture stated in the introduction.  At this time,
however, our understanding of the situation is still incomplete.  We
do not have a complete definition of string theory which allows us to
determine with certainty which low-energy models can or cannot be
realized within string theory.  It is also quite possible that string
theory imposes additional constraints, like the sign condition on the
gauge kinetic terms, which may be seen as quantum consistency
conditions given the data of the low-energy theory.  To prove or
disprove the conjecture of string universality in six dimensions, it
seems that further progress is needed in understanding both string
theory and low-energy supergravity.  There are also some purely
mathematical questions whose solutions may contribute significantly to
making progress in this direction.  In this concluding section, we
summarize some of the most relevant open questions, both
mathematical and physical, in this regard.
Most of these questions are discussed in further detail in \cite{tensors}.

\subsection{Mathematics questions}

There are a number of concrete mathematical questions related to
F-theory whose solutions would help clarify the precise set of
theories which can be realized in string theory via F-theory.
We briefly summarize here a few of these questions which are most
closely related to the discussion in  this paper.
\vspace*{0.05in}

\noindent
{\bf Weierstrass models}

We have argued that the correspondence between the low-energy data
associated with the anomaly cancellation conditions and the
mathematical structure of F-theory leads to a map (\ref{eq:map}) from
the anomaly lattice $\Lambda$ to the second homology lattice of the
F-theory base $B$.  In many cases this map associates with each nonabelian
factor in the gauge group an effective divisor
on $B$.  In order to show that the model can be realized in F-theory,
however, it is necessary to construct an explicit Weierstrass model
with the desired singularity locus associated with these divisors.
In many cases it seems that this can be done.  For a class of models
with $T = 1$ and gauge group $SU(N)$, we showed in \cite{KMT} that
there is a precise correspondence between the number of degrees of
freedom needed to fix the desired singularity structures on the proper
divisor locus and the number of degrees of freedom encoded in the
scalar matter fields through $H - V$.  It may be possible to give a
mathematical proof that the existence of a Weierstrass model follows
given some necessary topological conditions on the divisor locus.
Such a proof would be a useful step forward in identifying a large
class of the anomaly-free models which have F-theory realizations.
\vspace*{0.05in}

\noindent
{\bf Classifying matter singularities}

As mentioned above, while codimension one singularities in elliptic
fibrations are well-understood, codimension two singularities arising
at the intersection of codimension one singularities are not
completely understood.  Such singularities give rise in F-theory
compactifications to a variety of different representation structures
for matter fields.  In studying the set of anomaly-free supergravity
theories in six dimensions, we have encountered some novel matter
representations which may or may not be associated with valid F-theory
compactifications.  For example, in \cite{KMT}, we found some models
with an $SU(N)$ gauge group and matter in a
four-index antisymmetric representation.  While some F-theory
constructions are known which give rise to three-index antisymmetric
representations of $SU(N)$
\cite{Grassi-Morrison-2} there is as yet no understanding of a
singularity structure which would give a four-index antisymmetric
representation.  A complete classification of codimension two
singularities in elliptic fibrations of Calabi-Yau manifolds would be
of great assistance in systematically understanding the set of allowed
F-theory compactifications.
\vspace*{0.05in}

\noindent
{\bf Classification of elliptically fibered  Calabi-Yau manifolds}

It has been shown by Gross \cite{Gross} that the number of distinct
topological types of elliptically fibered Calabi-Yau threefolds is
finite up to birational equivalence.  Finiteness of the set of
topologically distinct elliptically-fibered Calabi-Yau threefolds is
shown in \cite{tensors} using minimal surface theory and the fact that
the Weierstrass form for an elliptic fibration over a fixed base has a
finite number of possible distinct singularity structures.  These
arguments, however, do not give a clear picture of how such
compactifications can be systematically classified.  A complete
mathematical classification of elliptically fibered Calabi-Yau
threefolds would be helpful in understanding the range of F-theory
compactifications.  The analogue of this question for four dimensions,
while probably much more difficult, would be of even greater interest,
since at this time we have very little handle on the scope of the
space of four dimensional supergravity theories which can be realized
through F-theory compactifications on Calabi-Yau fourfolds.

\subsection{Physics questions}

Some of the most difficult questions which need to be answered to
prove or disprove the conjecture of string universality in six
dimensions are essentially physical in nature.  The key questions
amount to

\noindent A) Are there as-yet unknown string compactifications which
extend the range of 6D theories beyond those attainable using known
F-theory constructions?

\noindent B) Can we identify new quantum consistency constraints on
low-energy theories which constrain the space of allowed models to
match more closely the set which can be realized through string
compactifications?
\vspace*{0.05in}

While A) is certainly an important question, it seems that the
most likely opportunity for narrowing the gap between the space of allowed
theories and the space of string constructions is by identifying new
constraints on low-energy theories.  If string universality is even
approximately correct, it is probably necessary to find a low-energy
quantum consistency condition which rules out the infinite families of theories
compatible with anomaly cancellation but which cannot be realized in
F-theory.  One place to look for such constraints is in the
set of conditions imposed by F-theory.  Considering the examples 3-5
in the previous section which cannot be realized in F-theory, we can
identify some possible constraints which may be needed in addition to
anomaly cancellation and the gauge kinetic term sign constraint.  It
seems plausible that some of these constraints may in fact be needed
for quantum consistency of any low-energy theory.  We briefly
mention here a few of these constraints, and make a few comments on
how they may be realized.
\vspace*{0.05in}

\noindent
{\bf  Unimodular lattice embedding}

As we see from example 3, a necessary condition for a given low-energy
theory to be realizable through F-theory is that the lattice $\Lambda$
associated with the anomaly cancellation constraints must be
embeddable in a unimodular lattice.  Associated with any of the
supersymmetric six dimensional gravity theories we are considering
here, there is a lattice of dyonic strings charged under the
anti-symmetric tensor fields of the theory, carrying a natural inner
product of signature $(1, T)$.  Some aspects of this lattice are
discussed in \cite{tensors}.  We do not know any reason why this
lattice must be unimodular for quantum consistency, but it is possible
that this may be necessary for unitary of the theory, or for some
other consistency reason.  Note that an analogous unimodularity
constraint follows from modular invariance for the heterotic
fundamental string construction.  Demonstrating that the string dyonic
lattice in a general 6D theory needs to be unimodular for a quantum
theory to exist would help in ruling out some of the apparently
consistent theories which we do not know how to realize in F-theory.

\vspace*{0.05in}

\noindent
{\bf Kodaira constraint and effective divisors}

As we noted in examples 4 and 5, some apparently consistent theories
cannot be realized in F-theory because the topological data for the
F-theory model realized through (\ref{eq:map}) leads to certain
divisors in F-theory not satisfying an effectiveness condition which
is needed for the F-theory construction.  In particular, the divisors
$-K, \xi_i$ associated with the vectors $-a, b_i$ must all be
effective (inside the Mori cone), as must the residual divisor locus
$Y$ associated with the vector $12a-\sum_{i}\nu_ib_i$.  This condition
has no obvious counterpart in the low-energy theory.  It seems
plausible, however, that some of these conditions may be associated
with sign constraints in the low-energy theory analogous to the sign
constraint on the gauge kinetic term.  For example, a term
proportional to $a \cdot j \;\tr R^2$ must appear in the action of the
theory by supersymmetry.  Arguments analogous to those in \cite{aa}
may suggest that this term must have a particular sign for consistency
with causality in field theory.  It is less clear how the condition on
the residual divisor locus can be seen from the low-energy theory, but
it seems possible that such a condition would arise from consistency
with supersymmetry, since this is the ultimate origin of the
Kodaira/Calabi-Yau condition in F-theory.
\vspace*{0.1in}

To conclude, we have not yet proven or disproven the conjecture of
string universality in six dimensions.  Simply pursuing this
conjecture, however, has given new insights into the global structure
of the space of 6D supergravity theories and string compactifications,
and has suggested some intriguing avenues for future progress, both in
physics, and in mathematics.

\bibliographystyle{amsplain}

\end{document}